\documentclass[journal]{IEEEtran}
\usepackage{graphicx}
\usepackage{epsf}
\usepackage{mathtools}
\usepackage{amssymb}
\usepackage{amsmath, amsfonts}
\usepackage{latexsym}
\usepackage{multirow}
\usepackage{multicol}
\usepackage[table]{xcolor}
\usepackage{color}

\usepackage{tabularx}
\usepackage{lipsum}

\usepackage{algorithm}
\usepackage{algorithmic}

\usepackage{mathrsfs}

\usepackage{fixltx2e}

\usepackage[mathscr]{euscript}

\usepackage{commath}
\usepackage{MnSymbol}

\usepackage{caption}
\usepackage{subcaption}

\usepackage{mathtools} 
\DeclarePairedDelimiterX\setc[2]{[}{]}{\,#1 \;\delimsize\vert\; #2\,}
\DeclarePairedDelimiterX\parth[2]{(}{)}{\,#1 \;\delimsize\vert\; #2\,}

\PassOptionsToPackage{hyphens}{url}
\usepackage[unicode=true, bookmarks=true, bookmarksnumbered=true, bookmarksopen=true, bookmarksopenlevel=1, breaklinks=false, pdfborder={0 0 0}, pdfborderstyle={}, backref=false, colorlinks=false]{hyperref}

\usepackage{theorem}
{
\theorembodyfont{\rmfamily}

}

\definecolor{orange}{RGB}{255,127,0}
\definecolor{blue}{RGB}{0,0,255}
\definecolor{red}{RGB}{255,0,0}
\definecolor{green}{RGB}{50,160,50}
\definecolor{grey}{RGB}{125,120,125}
\definecolor{purple}{RGB}{125,0,125}

\begin{document}
{
\title{{\fontsize{20}{2}\selectfont Dangerously Driven Cars Need to Go First}}

\author
{
Zachary Reyes, Seungmo Kim, \textit{Senior Member}, \textit{IEEE}, and Dhruba Sunuwar

\vspace{-0.3 in}

\thanks{Z. Reyes, S. Kim, and D. Sunuwar are with the Department of Electrical and Computer Engineering, Georgia Southern University, Statesboro, GA, USA. The corresponding author is S. Kim who can be reached at seungmokim@georgiasouthern.edu. This work is supported by the National Science Foundation (NSF) via award ECCS-2138446 and the Georgia Department of Transportation (GDOT) via grant RP21-08.}
}

\maketitle
\begin{abstract}
We propose that dangerously driven vehicles take a higher priority in multiple access for vehicle-to-everything communications (V2X). As more vehicles communicate, it is one's easy anticipation that the air interface will be crowded and thus a high magnitude of interference occurs. In response to this issue, we propose to prioritize the multiple access according to the driver's risky behavior while driving. Specifically, we build a driving simulator that aims at capturing the driver's distraction. The level of distraction will be measured in terms of (i) eye movement and (ii) motion. We build a number of different traffic scenarios including suburban highway, urban junction, etc. This research features an open source-based, thus low-cost, implementation of the driving simulator. Then, we apply the quantified driver's distraction level to the optimization of V2X multiple access.
\end{abstract}


\section{Introduction}\label{sec_intro}

\subsubsection{Background}
Connected vehicles (CVs) are no longer a futuristic dream seen in a science fiction, but they are emerging as reality in our everyday life. Throughout the short yet rich history of evolution of CVs, a salient paradigm is to accomplish the \textit{efficient data exchange} among vehicles \cite{cav_18}. It is now widely acknowledged that the safety of CVs heavily relies on the exchange of safety-critical messages \cite{access20}, which made vehicle-to-everything (V2X) communications a household name.

Despite the rapid evolution of V2X communications, successful delivery of the messages still remains an ambitious task mainly due to the high mobility and dynamicity of vehicular systems \cite{access19}. Furthermore, as more CVs are deployed, the number of exchanged messages will explode, which will likely deteriorate the successful message delivery performance \cite{federal}. As such, the \textit{priority} of the transmitted information needs to be carefully defined to ensure that CVs can obtain timely access to information that is important to them \cite{dave19}\cite{arxiv20}. It naturally translates to a particular question: \textit{based on what criteria the priority will be defined?}

To that line, this research designs a V2X system where a vehicle at a higher risk takes a higher priority since such a high-risk vehicle needs to inform its status to the neighboring vehicles more urgently for the road safety. However, it is a challenging task to precisely measure the driving risk of a vehicle due to (i) the complex interplay among diverse features defining the risk and (ii) their frequent change over time. This paper aims at addressing the challenge.

\subsubsection{Related Work}
Computer simulations were found efficient in experimenting out-of-vehicle features determining the driving risk already (e.g., weather, road condition, traffic condition, etc.) \cite{LeK98}\cite{DoR17}. This enables our simulator to focus on investigating \textit{in-vehicle} features, which define a driver's ability to safely operate a vehicle. In fact, previous research has found \textit{driver's distractions} and their correlation to accidents \cite{specees_70}. Many of the distractions outlined through this research are compatible with the pre-crash features identified by \cite{fars}, such as inattention (reduced cognitive thought), distractions by objects, and fatigue, have the potential for detection through the physiological measurement of eye movement.

Of various types, \textit{open-source simulators} particularly attract our interest. The Car Learning to Act (CARLA) is an open-source photorealistic simulator \cite{bib_carla_main} developed to support the development, training and validation of autonomous driving algorithms. It is written in C++ as a plugin of the Unreal Engine. The Simulation of Urban MObility (SUMO) \cite{bib_sumo} is another example, which features the ability to simulate a larger traffic network.

Several recent studies embodied realistic \textit{scenarios} based on such simulators. The CARLA Real Traffic Scenarios (CRTS) \cite{crts_arxiv21} is an example, which aims to provide a training and testing ground for autonomous driving systems. Another interesting example builds on the SUMO and provides highly interactive, artificial scenarios \cite{smarts}. A recent proposal replays accidents in the simulator as a means to prove that their autonomous driving system reacts more swiftly than human drivers can \cite{waymo}.

\subsubsection{Contributions}
Building on the state-of-the-art, this research aims at improvement on the following two fronts:
\begin{itemize}
\item Design a simulator dedicated to \textit{distracted driving at a low cost}
\item \textit{Optimize V2X communications} based on the driving risk-based prioritization of safety messages
\end{itemize}

\section{System Model}\label{sec_model}
The simulator will be composed of three main software components: (i) RoadRunner for map generation and traffic signalization; (ii) SUMO for vehicle and pedestrian routing; and (iii) CARLA for simulation and real-time analysis via a modified version of Unreal Engine 4 (UE4). Fig. \ref{fig_simulator_overview_sw} shows the overall structure of the simulator. Each component will be discussed further in this section.

\begin{figure*}
\centering
\begin{subfigure}[b]{0.8\textwidth}
\centering
\includegraphics[width = \linewidth]{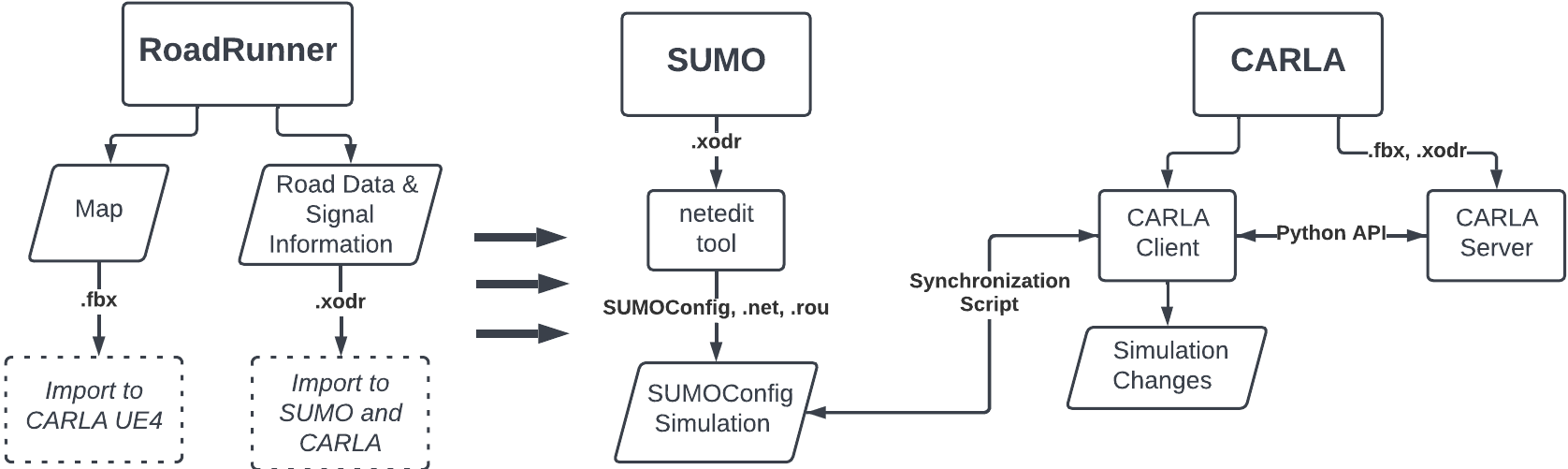}
\caption{Implementation components}
\label{fig_simulator_overview_sw}
\end{subfigure}
\begin{subfigure}[b]{0.8\textwidth}
\vspace{0.2 in}
\centering
\includegraphics[width = \linewidth]{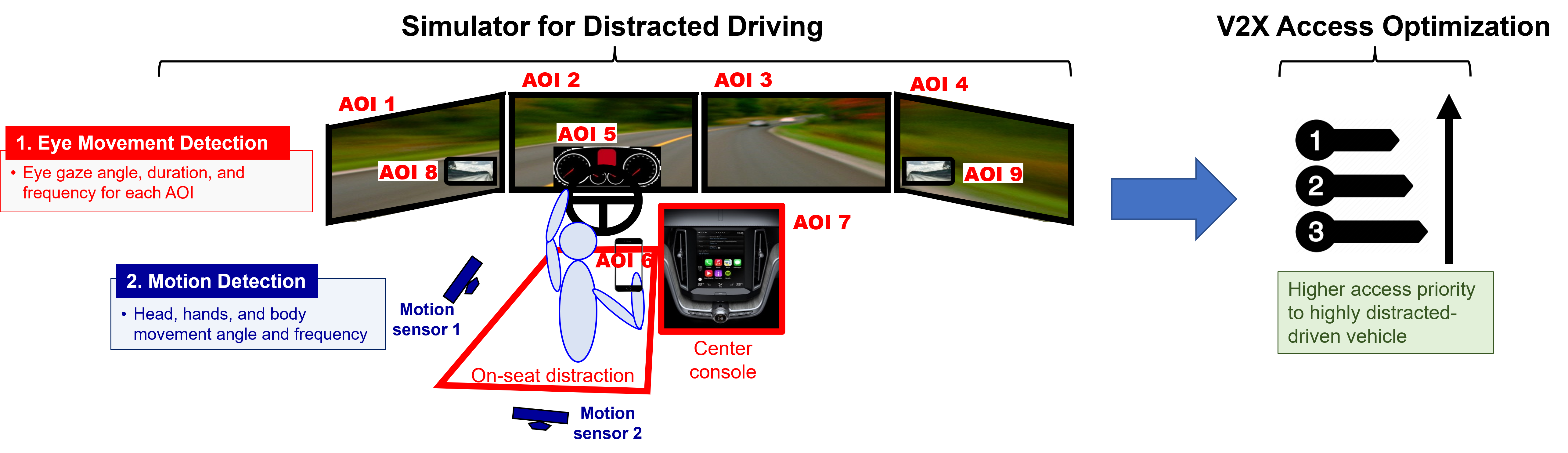}
\caption{Driving risk-adaptive V2X optimization}
\label{fig_simulator_overview_v2x}
\end{subfigure}
\caption{Overview of the proposed driving simulator}
\label{fig_simulator_overview}
\end{figure*}

\subsection{RoadRunner}
RoadRunner is an interactive editor that allows for three-dimensional (3D) scene design for simulating and testing driving systems \cite{bib_roadrunner}. The proposed simulator utilizes RoadRunner to create a realistic scenario to describe normal driving situations such as roads, road lines, sidewalks, stoplights, and signs, as well as other objects a driver may see while driving, such as poles, trees, and buildings. The 3D map terrain (.fbx), road data and signal information (.xodr), is generated and exported to SUMO for simulation configuration, and CARLA for realistic visualization. 

\subsection{Simulation of Urban MObility (SUMO)} 
How a person responds to certain traffic situations is an important part of analyzing driving behavior. SUMO is an open-source multi-modal traffic simulation package designed to handle large networks \cite{bib_sumo}, and is utilized in the simulator for defining repeatable traffic scenarios that a user can co-simulate and interact with via CARLA.

NETEDIT is the visual tool within SUMO to edit networks generated by the exported OpenDRIVE (.xodr) file from RoadRunner. Routes are created to give vehicles waypoints to drive toward, respecting road laws such as stop lights, signs, and yielding. Pedestrians are also able to be given routes, to provide more realism in the scenario. 

It is also noteworthy, though, that we consider retiring SUMO as the main traffic definition and scenario creation tool due to its lack of support for complex junctions, as well as lack of CARLA application programming interface (API) support for SUMO spawned vehicles.

\subsection{Car Learning to Act (CARLA)} 
CARLA features modular and flexible APIs to address issues in driving related research. CARLA is based on UE4 \cite{bib_UE4} to run simulations and uses the OpenDRIVE standard to define roads and urban settings, as created in RoadRunner. Control over the simulation can be accessed real-time through the API handled in Python and C++ \cite{bib_carla_main}.

This simulation adopts CARLA v0.9.13 for implementation of realistic 3D visualization to the 2D simulation of SUMO via Co-Simulation. Vehicles and pedestrians are spawned in CARLA as the SUMO simulation begins, allowing the main user to interact with the world through the manually driven ego vehicle, spawned and managed by the Python API.

For traffic generation of the proposed simulator, we deem the value of the CARLA Traffic Manager \cite{bib_carla_trafficmanager} higher than that of SUMO in the sense that the former allows full access to vehicles even in the middle of a simulation running.

\section{Proposed Implementation Mechanism}\label{sec_proposed}
\subsection{Driving Simulator Implementation}
Fig. \ref{fig_simulator_overview} demonstrates a zoomed-out look of the entire proposed driving simulator. In particular, Fig. \ref{fig_simulator_overview_v2x} depicts a setup for the proposed driving simulator and its connection to the V2X access optimization.

To elaborate Fig. \ref{fig_simulator_overview_v2x}, in driving circumstances, there are numerous factors that determine the \textit{distraction level of a driver}. We specify that the proposed simulator will focus on the measurement of (i) eye movements and (ii) motions of head, hands, and body while driving. The simulator consists of a main workstation computer connected to 4 monitors, a steering wheel, and a pedal system to actualize the participant's driving experience. The level of \textit{motion} distraction will be measured two motion sensors near the driver for head, hand, and body movement in terms of the intensity and frequency. Also, the participant will be wearing eye-tracking glasses to measure the eye gaze angle, duration, and frequency for each area of interest (AOI), which is specified in Fig. \ref{fig_simulator_overview_v2x}.

The initial driving scenario will be composed of a map built on RoadRunner, which Fig. \ref{fig_RR_map} depicts. The RoadRunner features the ability to add 3D assets placed in UE4 to provide realism to the so-called ``world'' as shown in Fig. \ref{fig_UE4_map_birdseye}. The world can be added with further details such as stop signs, yield signs, a roundabout, and a four-way intersection handled by stop lights, as shown in Figs. \ref{fig_stopsign} and \ref{fig_fourway_stoplights}.

\begin{figure}
\centering
\begin{subfigure}[b]{0.5\textwidth}
\centering
\includegraphics[width = \linewidth, height = 1.8 in]{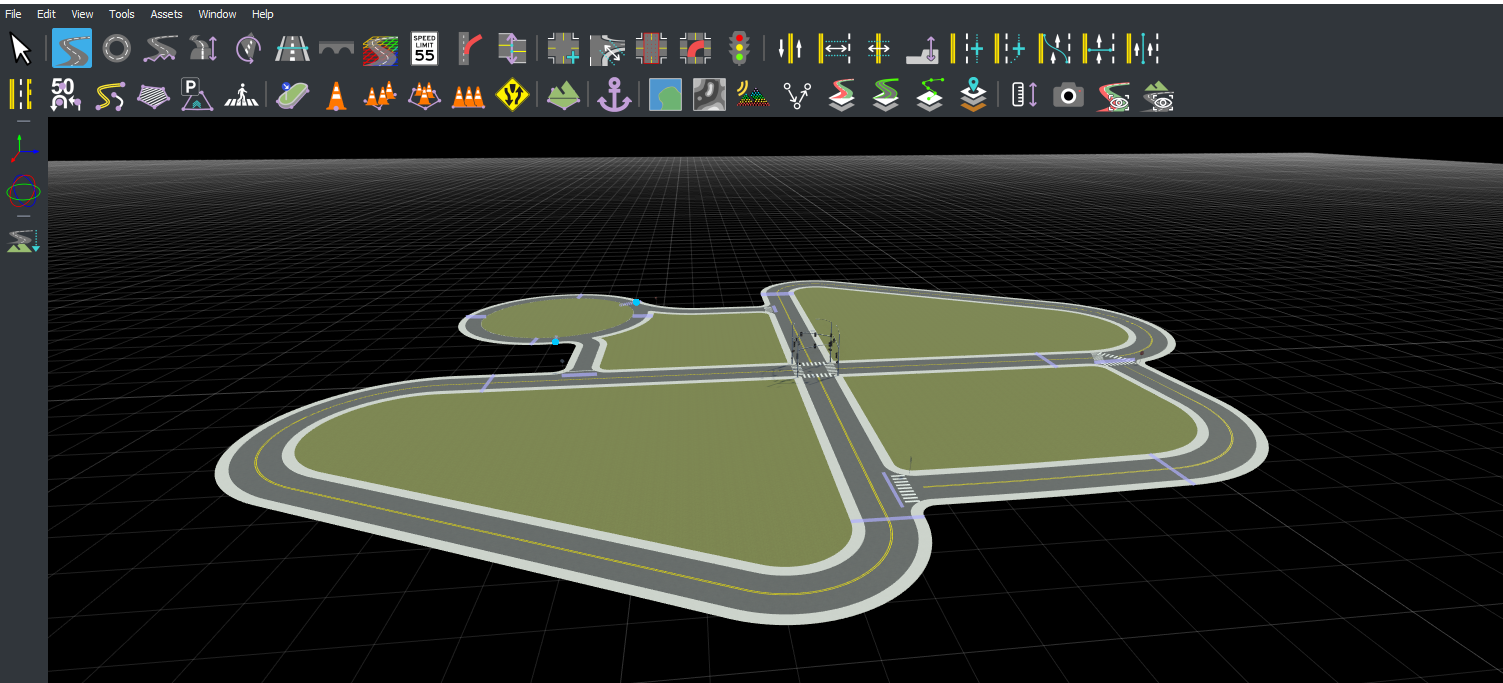}
\caption{On RoadRunner}
\label{fig_RR_map}
\end{subfigure}
\begin{subfigure}[b]{0.5\textwidth}
\vspace{0.1 in}
\centering
\includegraphics[width = \linewidth, height = 1.8 in]{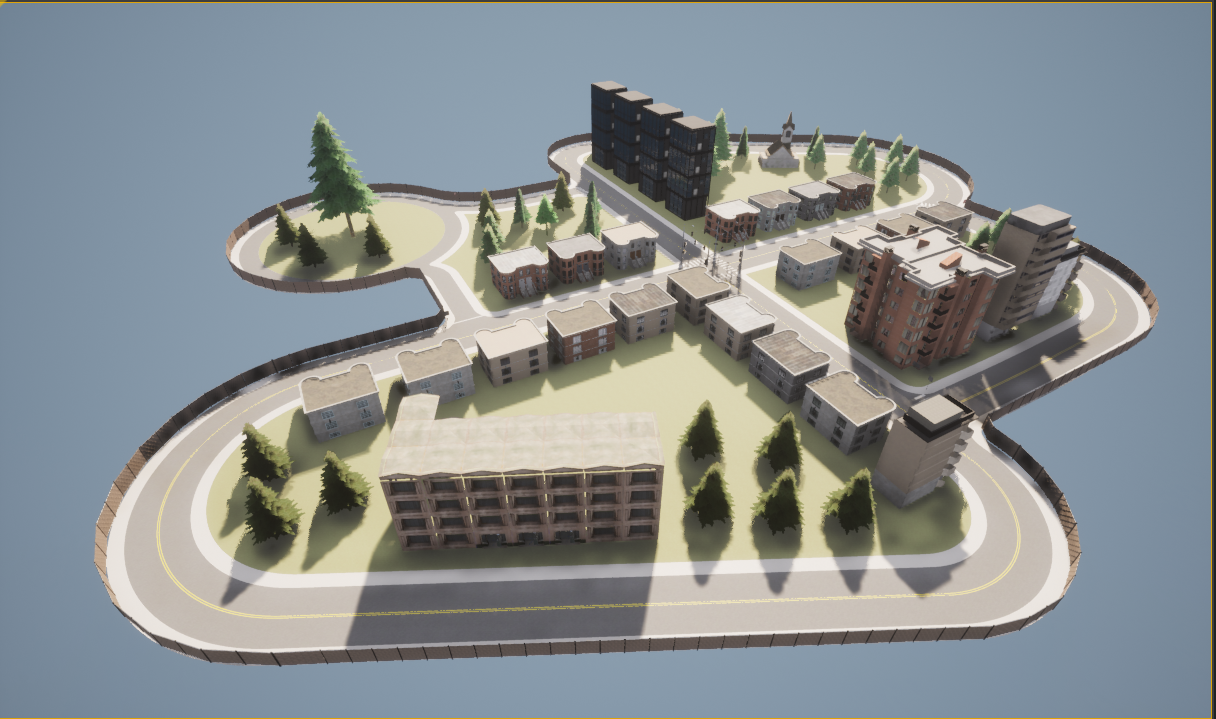}
\caption{On UE4}
\label{fig_UE4_map_birdseye}
\end{subfigure}
\caption{Initial road layout}
\end{figure}

\begin{figure}
\centering
\begin{subfigure}[b]{0.5\textwidth}
\centering
\includegraphics[width = \linewidth, height = 1.8 in]{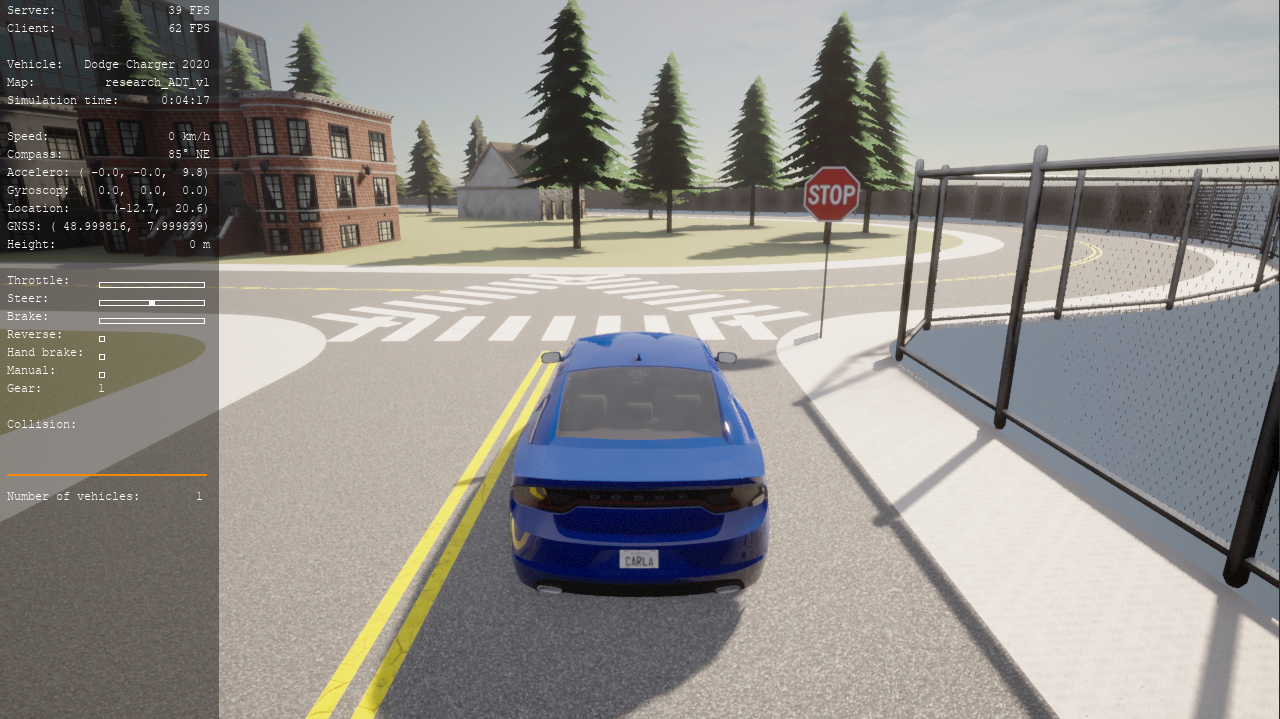}
\caption{Stop sign road condition}
\label{fig_stopsign}
\end{subfigure}
\begin{subfigure}[b]{0.5\textwidth}
\vspace{0.1 in}
\centering
\includegraphics[width = \linewidth, height = 1.8 in]{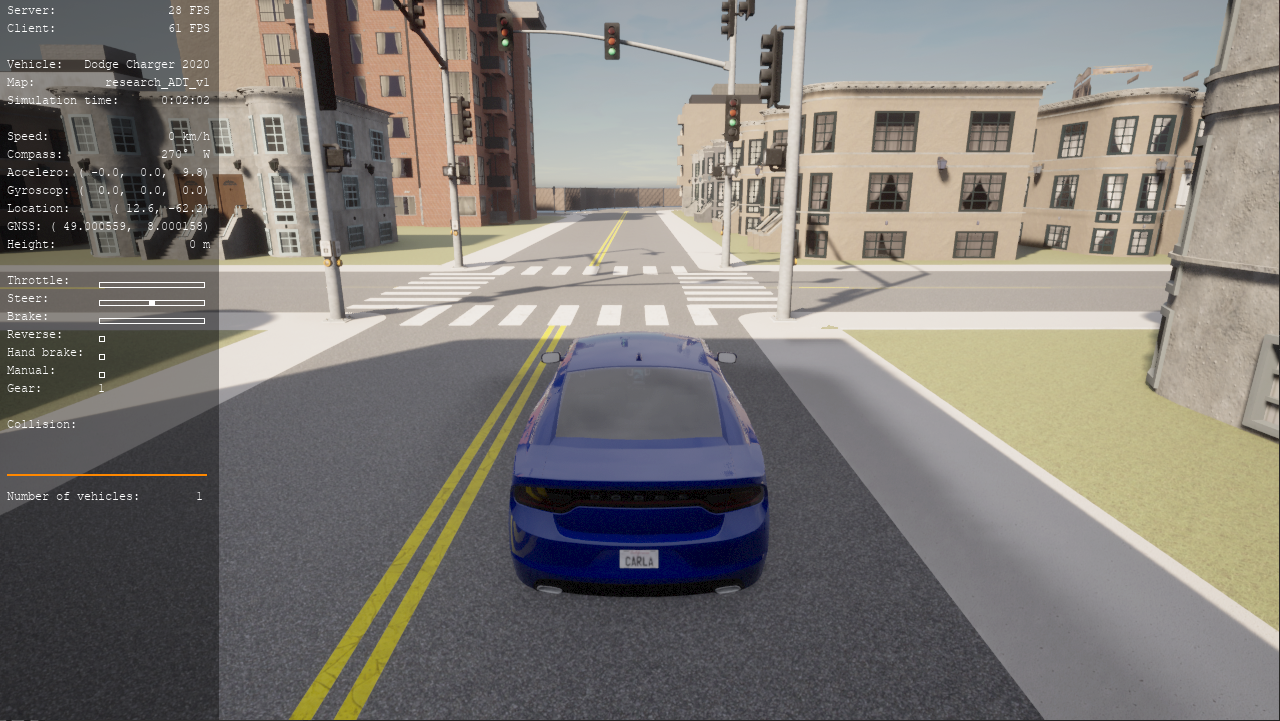}
\caption{Four-way intersection stop lights}
\label{fig_fourway_stoplights}
\end{subfigure}
\caption{Adding details of traffic scenarios on CARLA}
\end{figure}

We also implement the first-person view that enables manual driving of the vehicle for a higher level of immersion, which Fig. \ref{fig_firstPerson_interacting} shows. This implementation has already been applied to our V2X access simulation \cite{arxiv2302}, where how the successful delivery of a safety message can actually improve the driver's experience.

\begin{figure} 
\centering
\includegraphics[width = \linewidth, height = 1.8 in]{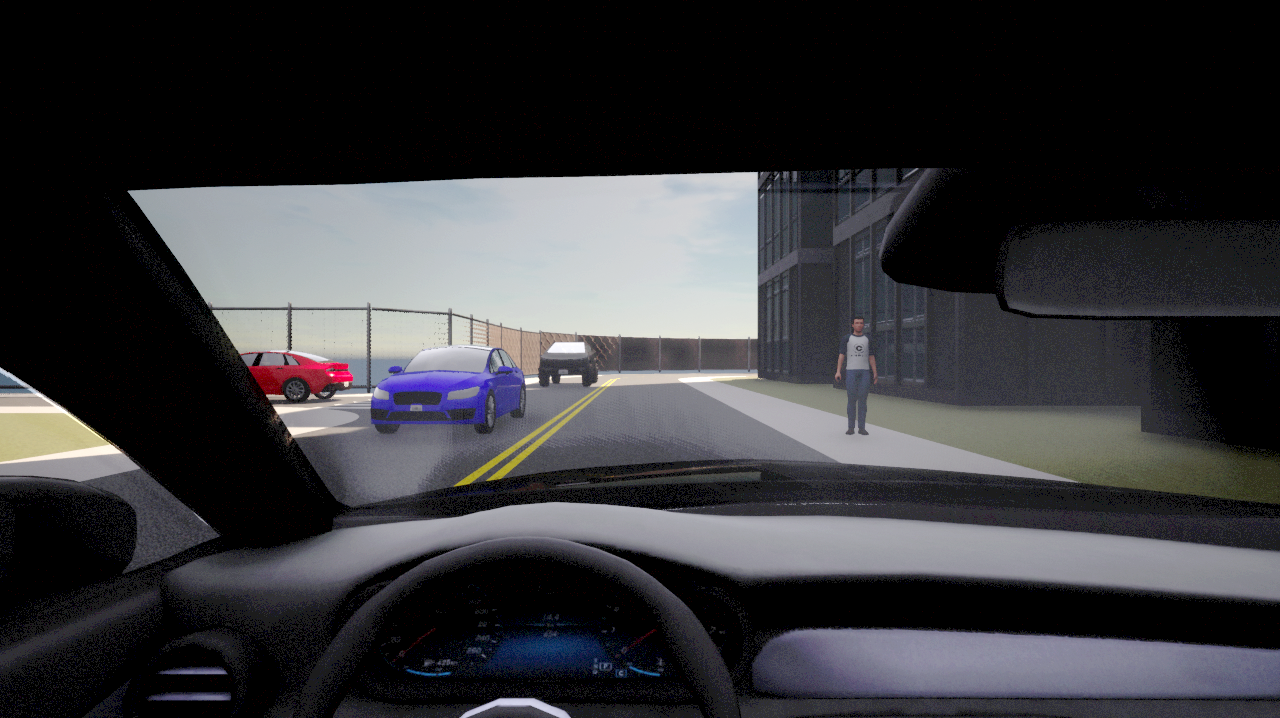}
\caption{First-person view of an urban scenario on CARLA}
\label{fig_firstPerson_interacting}
\end{figure}

\begin{figure}[h]
\vspace{-0.17 in}
\centering
\includegraphics[width = 0.8\linewidth]{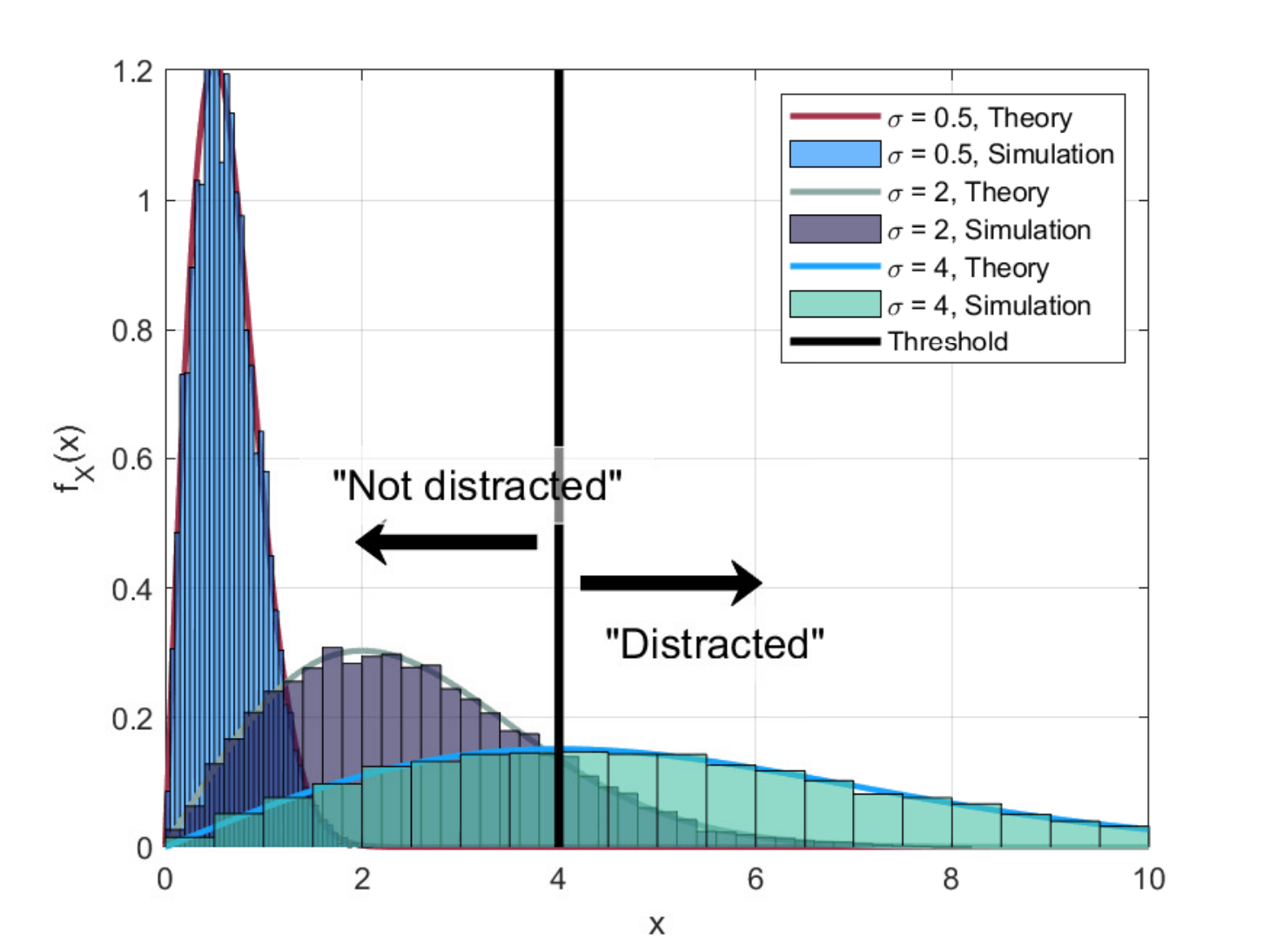}
\caption{Probabilistic characterization of drivers' distraction}
\label{fig_rayleigh}
\end{figure}

\subsection{Quantification of Drivers' Distraction}\label{sec_proposed_rayleigh}
Fig. \ref{fig_rayleigh} displays the distribution of driver's distraction level on an arbitrary road defined as a two-dimensional Borel set $\mathbb{R}^2$. We propose to model the distraction level as a Rayleigh distribution $f_{X}(x) = \frac{x}{\sigma^2} e^{\frac{-x^2}{2\sigma^2}}, \hspace{0.01 in} x \ge 0$ \cite{discrete_rayleigh_04}, where the continuous random variable $x$ indicates the \textit{level of drivers' distraction among all the vehicles} in a certain borel set $\mathbb{R}^2$. As such, the parameter $\sigma$ is set to indicate the drivers \textit{mean distraction level} among the vehicles with an appropriate scaling, i.e., $\mathbb{E}\left[x\right] = \frac{\pi \sigma}{2}$.

The rationale of this modeling is as follows. The assumption of $x$ to be a continuous random variable is plausible in the sense that the distraction level will naturally be continuously quantified especially when measured among numerous ensembles (i.e., different drivers). In other words, no two different drivers can be mapped to the same distraction level even with the same type of distracting behavior (e.g., the same angle and frequency of eye movement off the windshield), considering the drivers' different profiles (e.g., driving experience, surrounding traffic, etc.). Now, here is how the random variable $x$ is quantified. As Fig. \ref{fig_rayleigh} presents, we define $x=0$ to be the \textit{perfectly concentrated} state and $x > 0$ to be a \textit{distracted} state. Given that, it is straightforward that $x \rightarrow \infty$ indicates the distraction level getting elevated. As such, it is a reasonable assumption that on a normal road, the peak is located more toward the \textit{not distracted} side as $\lambda$ is with a larger value.

For generation of Fig. \ref{fig_rayleigh}, we created 100 drivers on a sample 1 km-by-1 km, 2-junction urban model, and assigned random distraction levels $x$ following the $f_{X}(x)$ defined as a Rayleigh distribution. We mapped this simulation to the probability density function (PDF) $f_{X}(x)$, thereby confirming the statistical validity of the simulation.

Now, we set a threshold, $\theta$, on a certain environment that is represented by a $\sigma$. This determines a certain proportion of the drivers to be claimed as ``distracted drivers,'' to which the highest priority of V2X transmission is assigned. As such, the proportion of distracted drivers on the arbitrary road environment $\mathbb{R}^2$ can be found as $\mathbb{P}(X \ge \theta) = \int_{\theta}^{\infty} f_{X}(x) \text{d}x$. For instance, with $\sigma = 2$, the proportion can be found as
\begin{align}\label{eq_thresholds}
\mathbb{P}(X \ge \theta) = \begin{cases} 3.3546 \times 10^{-4}, \hspace{0.1 in} \theta = 8\\
0.1353, \hspace{0.5 in} \theta = 4\\
0.6065, \hspace{0.5 in} \theta = 2\end{cases}
\end{align}
We reiterate that a higher level of $\theta$ indicates a ``stricter'' definition of distracted driving, so fewer drivers can be prioritized in V2X transmission. Simulation results on this shall follow in Section \ref{sec_results_v2x}.

\section{Simulation Results}\label{sec_results}

\subsection{Driving Simulator Runtime Verification}\label{sec_results_runtime}
Each scenario must be tested and verified to be reproducible for each participant. This process is done through setting vehicle and pedestrian routes, in terms of pathing and timing, to be equal within the scenario for each reset of the simulation. This section will discuss the three methods of routing tested, followed by the method the proposed simulator utilizes. 

\subsubsection{SUMO}
SUMO was the initial method for setting and testing vehicle routes, as CARLA supports SUMO co-simulation. The NETEDIT tool allows for simple vehicle routing and placement over connected roads between junctions as shown by the yellow line, indicating a vehicle route, in Fig. \ref{fig_SUMO_routes}. While this worked well for vehicle placement, the end result of the simulation in CARLA was not realistic due to lack of support in SUMO for complex junctions, which can be seen in the same figure.

\subsubsection{RoadRunner Scenario}
The RoadRunner Scenario (the ``RR Scenario'' hereafter) simulation editor allows for design of scenarios within scenes by placing vehicles and paths, defining logic, and setting parameters for scenarios \cite{bib_roadrunner_scenario} which can then be simulated in the editor or exported to CARLA.

It supports CARLA co-simulation by placing vehicles, defining waypoints, and setting timing logic in RR Scenario. These vehicles are then given behaviors via the Python API to be defined in CARLA, allowing CARLA to have control of them rather than RR Scenario. An example scenario can be seen in Fig. \ref{fig_RRscenario}. Each vehicle in the scenario is spawned by RR Scenario, but controlled by CARLA via the \texttt{CARLA.rrbehavior} attribute, tying each vehicle to a defined Python behavior script. When running the simulation, what is shown in the RR Scenario Simulation happens in CARLA, and can be interacted by the participant via the Python API.

\begin{figure}
\begin{subfigure}[b]{\linewidth}
\centering
\includegraphics[width = 0.9\linewidth, height = 1.8 in]{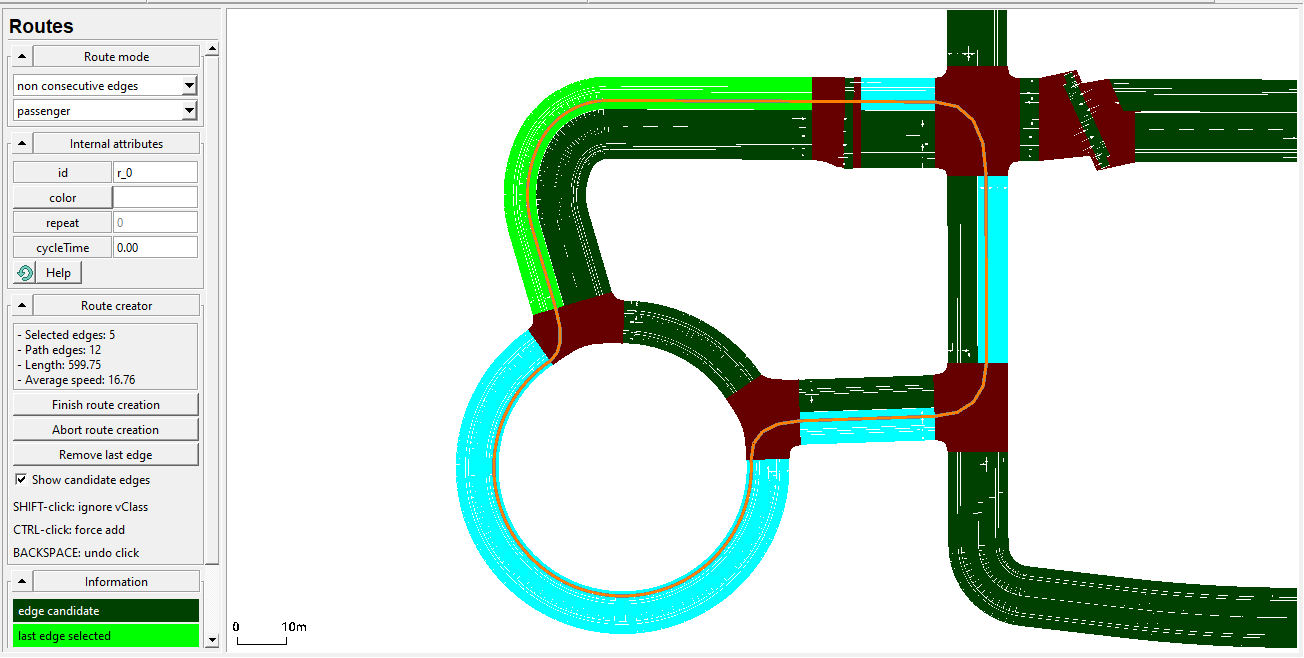}
\caption{Routing example on SUMO NETEDIT}
\label{fig_SUMO_routes}
\end{subfigure}
\begin{subfigure}[b]{\linewidth}
\centering
\includegraphics[width = 0.9\linewidth, height = 1.8 in]{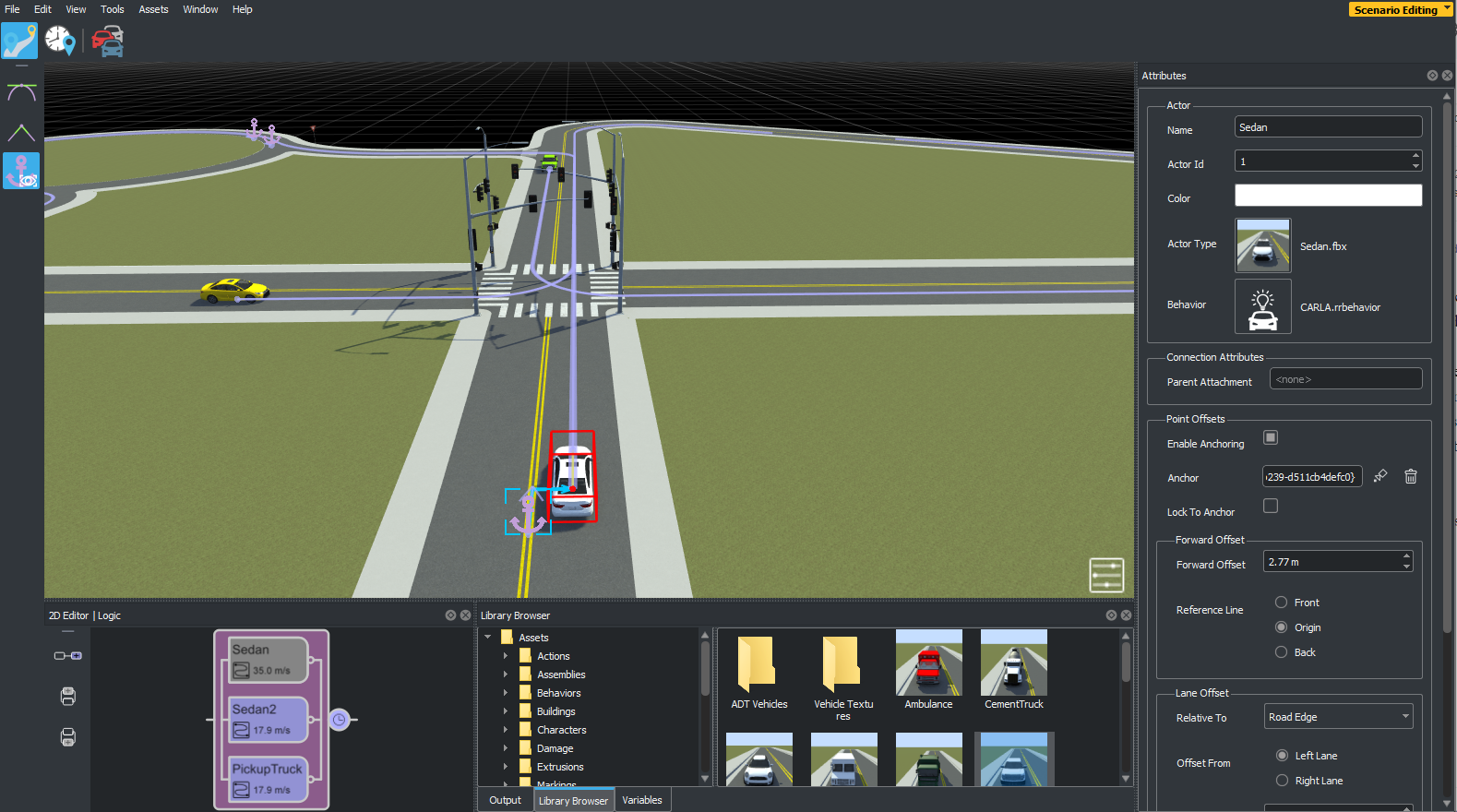}
\caption{RoadRunner Scenario simulation with CARLA vehicle behaviors}
\label{fig_RRscenario}
\end{subfigure}
\begin{subfigure}[b]{\linewidth}
\centering
\includegraphics[width = 0.9\linewidth, height = 1.8 in]{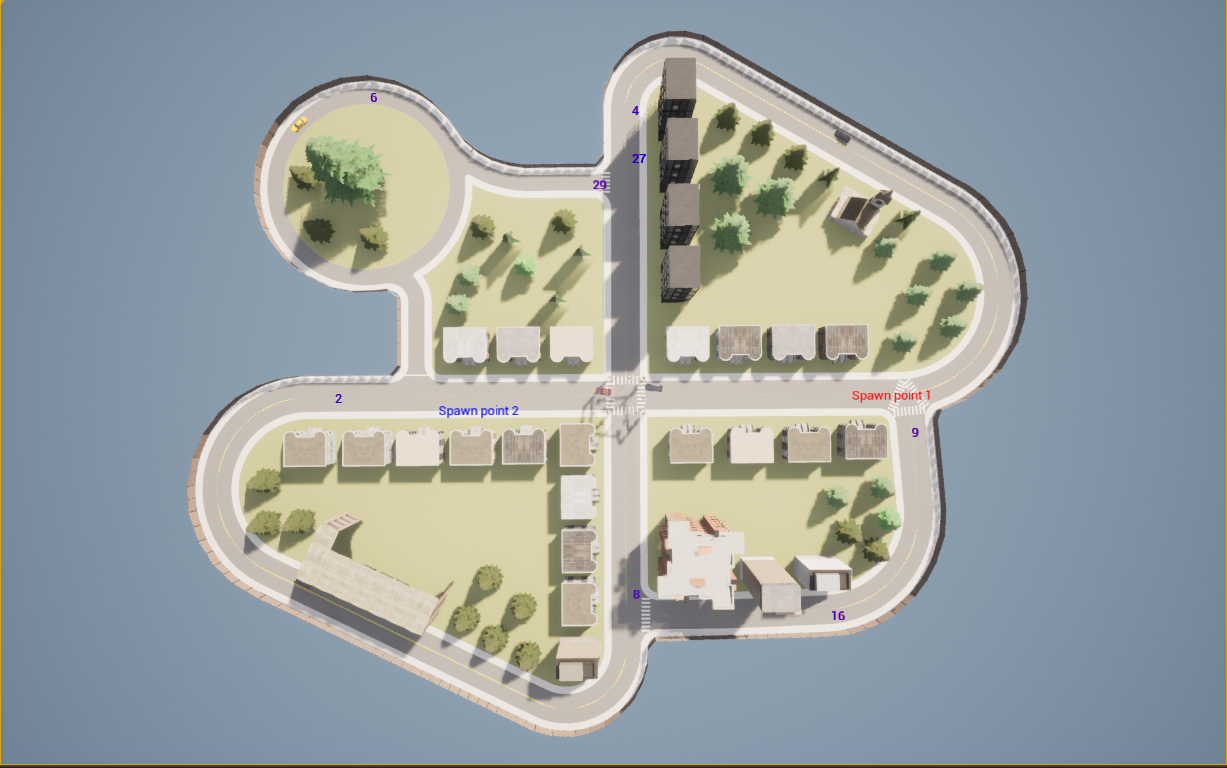}
\caption{CARLA spawn points for vehicle routing}
\label{fig_scenario_vehicle_waypoints}
\end{subfigure}
\caption{Implementation of specific scenarios}
\label{fig_scenarios}
\end{figure}

\begin{figure*}
\centering
\begin{subfigure}[b]{0.32\textwidth}
\hfill
\centering
\includegraphics[width = \linewidth]{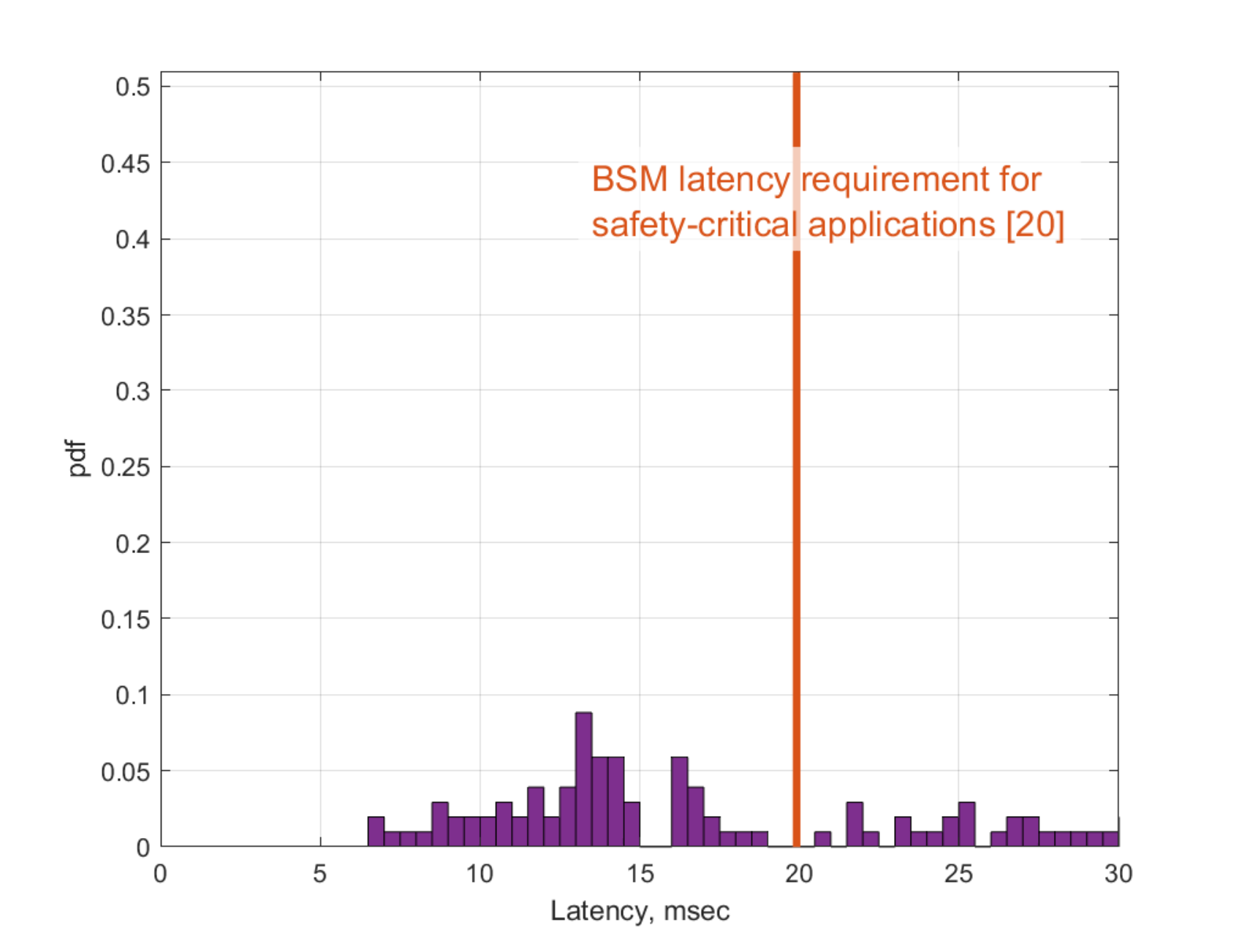}
\caption{Loosely prioritized ($\theta = 2$)}
\label{fig_latency_lambda17_theta2}
\end{subfigure}
\begin{subfigure}[b]{0.32\textwidth}
\hfill
\centering
\includegraphics[width = \linewidth]{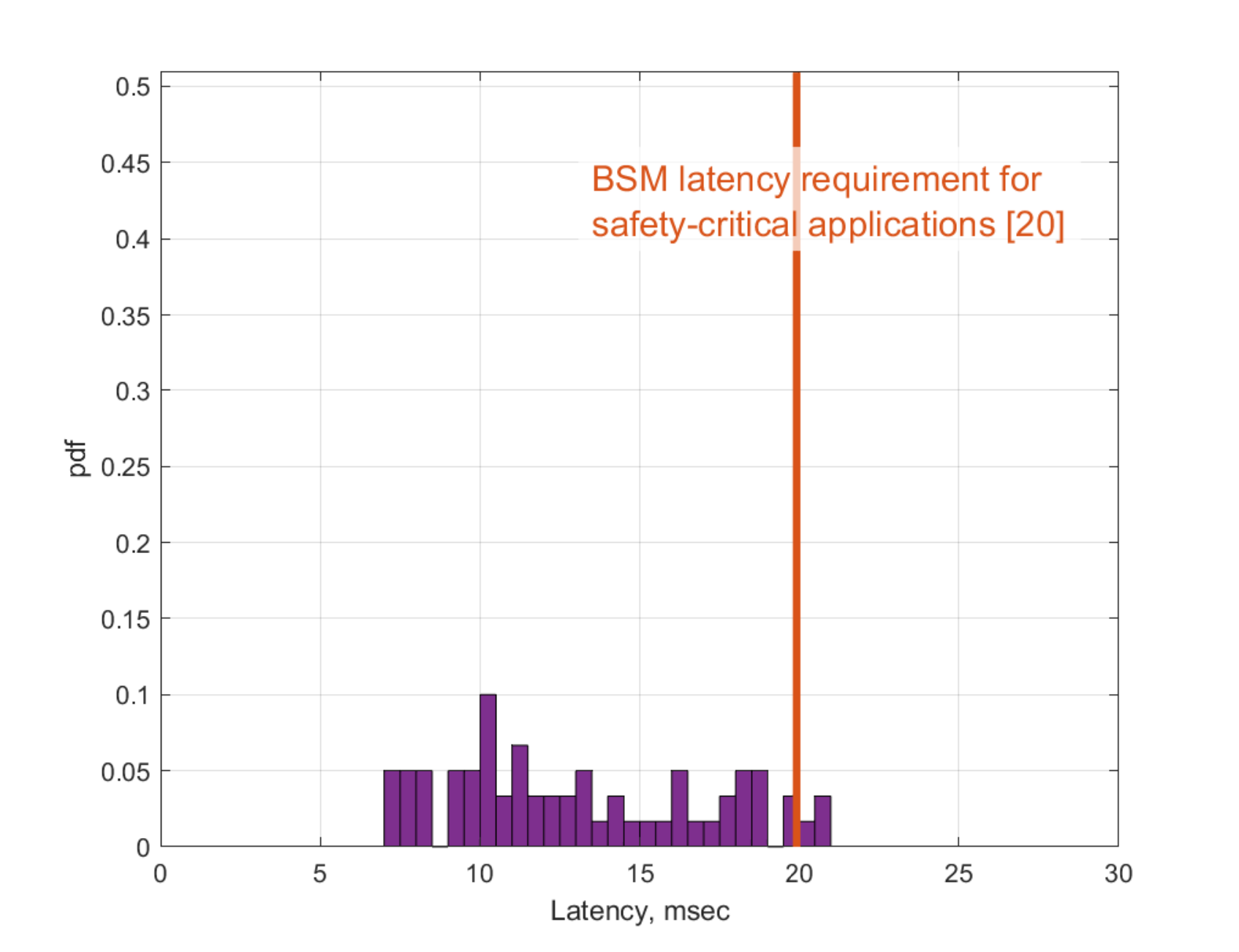}
\caption{Medium ($\theta = 4$)}
\label{fig_latency_lambda10_thetq4}
\end{subfigure}
\begin{subfigure}[b]{0.32\textwidth}
\centering
\includegraphics[width = \linewidth]{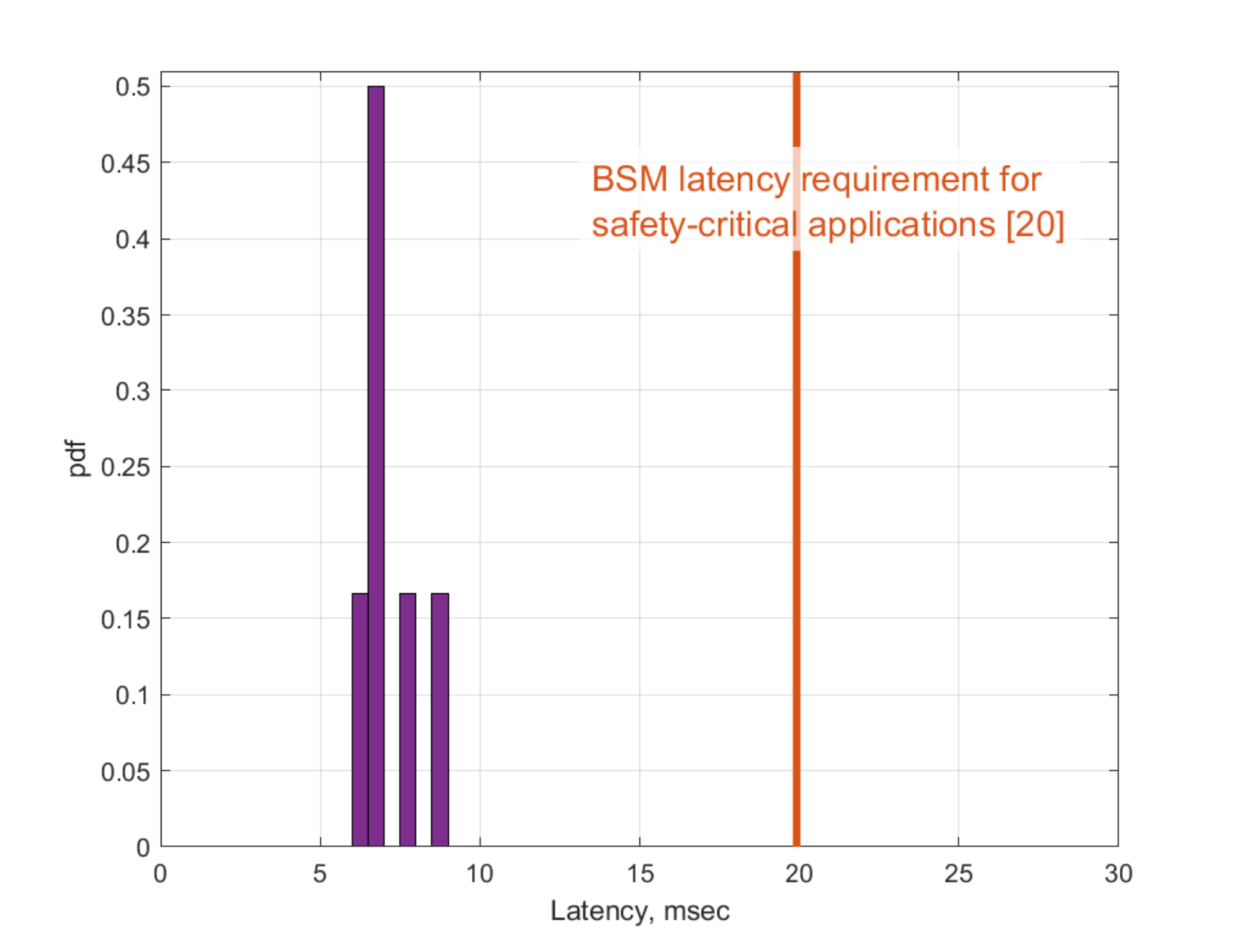}
\caption{Strictly prioritized ($\theta = 8$)}
\label{fig_latency_lambda1_theta8}
\end{subfigure}
\caption{Distribution of C-V2X end-to-end latency according to the threshold on ``distraction,'' $\theta$}
\label{fig_latency}
\end{figure*}

The only flaw of using RR Scenario for co-simulation is the lack of support for traffic signals. Though the signalization logic for lights and signs are created when designing the scene, the RoadRunner placed actors do not react to real-time traffic laws such as a stop light or yield sign, even if their behavior script defines them to do so. The actors follow their path with no regard to the world around them, making this unsatisfactory for simulation. If in the future this can be changed, RR Scenario would be the optimal tool for simulating these scenarios.

\subsubsection{CARLA}
To use CARLA for generating scenarios requires using the Python API to generate waypoints for each vehicle and pedestrian to navigate to. The waypoints for each vehicle can be defined by a list of respective UE4 vehicle spawn points, which are gathered by a Python script. Fig. \ref{fig_scenario_vehicle_waypoints} shows a diagram of possible spawn points for vehicles to navigate to and from, acting as a route for each vehicle.

Each actor is then given instructions to follow through the CARLA Traffic Manager, which can control actor behavior, such as ignoring lights or stop signs. For the initial scenario, all vehicles and pedestrians will follow all traffic laws to provide a baseline for future scenario simulations.


\subsection{Application to V2X Optimization}\label{sec_results_v2x}
Fig. \ref{fig_latency} illustrates the statistics that was described in Section \ref{sec_proposed_rayleigh}. That is, the latency of a C-V2X system improves as the parameter $\lambda$ is increased. For this simulation, we implemented an urban microcell (UMi) environment defined in the 3rd Generation Partnership Project (3GPP) Technical Realization (TR) 38.901 \cite{tr38901}. We built a separate simulator measuring the performance of a New Radio (NR)-V2X system based on the 3GPP Release 17 physical and radio resource control (RRC) standards, e.g., Technical Specification (TS) 38.331 \cite{ts38331}. We deployed 100 vehicles on a 2-junction road environment, the priority of transmissions is given to the distraction level exceeding thresholds $\theta = \{2,4,8\}$ as was provided in Eq. (\ref{eq_thresholds}). As the Figs. \ref{fig_latency_lambda17_theta2} through \ref{fig_latency_lambda1_theta8} show, a more strictly prioritized traffic yields a lower latency.

Notice that the most critical applications require the basic safety message (BSM) with the latency no greater than 20 msec \cite{arxiv2302}. This implies that the chance of the applications being successfully supported gets higher as the vehicles are more strictly prioritized from Fig. \ref{fig_latency_lambda17_theta2} to \ref{fig_latency_lambda1_theta8}.

\section{Conclusions}\label{sec_conclusions}
This paper proposed to prioritize dangerously driven vehicles in multiple access in a V2X system. It also proposed to define the driving risk based on the driver's distraction level. As an effort to actualize the idea, we built a driving simulator and a number of representative traffic scenarios via an open source-based (hence low-cost) implementation. The driver's distraction level quantified as such was used as the key variable in the optimization of V2X multiple access. Our simulation results proved that the proposed driving danger-based V2X access prioritization method is effective in improving the performance of a V2X system, which was measured in the end-to-end latency.



\begin{thebibliography}{99}
\setlength{\parskip}{0.0000001 em}

\bibitem{cav_18} M. Islam, M. Rahman, S. M. Khan, and M. Chowdhury, ``Connected vehicle application development platform (CVDeP) for edge-centric cyberphysical systems,'' \textit{arXiv:1812.11648}, Dec. 2018.

\bibitem{access20} S. Kim and B. J. Kim, ``Crash risk-based prioritization of basic safety message in DSRC,'' \textit{IEEE Access}, vol. 8, Nov. 2020.

\bibitem{access19} S. Kim, ``Impacts of mobility on performance of blockchain in VANET,'' \textit{IEEE Access}, vol. 7, May 2019.

\bibitem{federal} U.S. FCC, ``Use of the 5.850-5.925 GHz band,'' \textit{Federal Register}, vol. 86, no. 83, May 2021.

\bibitem{dave19} T. Dessalgn and S. Kim, ``Danger aware vehicular networking,'' in \textit{Proc. IEEE SoutheastCon 2019}.

\bibitem{arxiv20} S. Kim and B. J. Kim, ``Novel backoff mechanism for mitigation of congestion in DSRC broadcast,'' \textit{arXiv:2005.08921}, May 2020.

\bibitem{LeK98} W. S. Lee, J. H. Kim, and J. H. Cho, ``A driving simulator as a virtual reality tool,'' in \textit{Proc. IEEE International Conference on Robotics and Automation 1998}.

\bibitem{DoR17} A. Dosovitskiy, G. Ros, F. Codevilla, A. Lopez, and V. Koltun, ``CARLA: An open urban driving simulator,'' in \textit{Proc. Conference on Robot Learning 2017}.

\bibitem{specees_70} K. Young and P. Salmon, ``Examining the relationship between driver distraction and driving errors: A discussion of theory, studies and methods,'' \textit{Safety Science}, vol. 50, no. 2, Feb. 2012.

\bibitem{fars} Website of National Highway Traffic Safety Administration (NHTSA) Fatality Analysis Reporting System (FARS) database, [Online]. Available: \url{https://www-fars.nhtsa.dot.gov//QueryTool/QuerySection/SelectYear.aspx}

\bibitem{bib_carla_main} ``CARLA simulator,'' \textit{carla.readthedocs.io.}, [Online]. Available: \url{https://carla.readthedocs.io/en/0.9.13/}
‌
\bibitem{bib_sumo} Website of Eclipse SUMO - Simulation of Urban MObility, [Online]. Available: \url{https://www.eclipse.org/sumo/}

\bibitem{crts_arxiv21} B. Osinski, P. Milos, A. Jakubowski, P. Ziecina, M. Martyniak, C.Galias, A. Breuer, S. Homoceanu, and H. Michalewski, ``CARLA real traffic scenarios--novel training ground and benchmark for autonomous driving,'' \textit{arXiv preprint arXiv:2012.11329}, Dec. 2020.

\bibitem{smarts} M. Zhou, J. Luo, J. Villella, Y. Yang, D. Rusu, J. Miao, W. Zhang,
M. Alban, I. Fadakar, Z. Chen, and A. Huang, ``Smarts: Scalable multi-agent reinforcement learning training school for autonomous driving,'' \textit{arXiv preprint arXiv:2010.09776}, Oct. 2020.

\bibitem{waymo} J. M. Scanlon, K. D. Kusano, T. Daniel, C. Alderson, A. Ogle, and T. Victor, ``Waymo simulated driving behavior in reconstructed fatal crashes within an autonomous vehicle operating domain,'' \textit{Accident Anal. Prevention}, vol. 163, Dec. 2021.


\bibitem{bib_roadrunner} Mathworks, ``Design 3D scenes for automated driving simulation,'' [Online]. Available: \url{https://www.mathworks.com/products/roadrunner.html}

\bibitem{bib_carla_trafficmanager} ``Traffic Manager - CARLA Simulator,'' [Online]. Available: \url{https://carla.readthedocs.io/en/0.9.13/adv_traffic_manager/}

\bibitem{bib_UE4} Unreal Engine, ``The most powerful real-time 3D creation tool,'' [Online]. Available: \url{https://www.unrealengine.com/en-US}

\bibitem{bib_roadrunner_scenario} Mathworks, ``RoadRunner scenario,'' [Online]. Available: \url{https://www.mathworks.com/products/roadrunner-scenario.html}


\bibitem{arxiv2302} D. Sunuwar, S. Kim, and Z. Reyes, ``Is 30 MHz enough for C-V2X?,'' \textit{arXiv:2302.09536}, Feb. 2023.

\bibitem{discrete_rayleigh_04} M. Siddiqui, ``Statistical inference for Rayleigh distributions,'' \textit{J. Res. Nat. Bureau Standards, Sec. D: Radio Sci.}, vol. 68D, no. 9, p. 1007, 1964.

\bibitem{tr38901} 3GPP, ``5G; Study on channel model for frequencies from 0.5 to 100 GHz,'' \textit{ETSI TR 138901}, V17.0.0, Apr. 2022.

\bibitem{ts38331} 3GPP, ``5G; NR; Radio resource control (RRC); Protocol specification,'' \textit{ETSI TS 138331}. V17.3.0, Jan. 2023.

\end{thebibliography}
\end{document}